\documentstyle{article}
\begin{document}

~\hfill{\tt hep-th/9908028}

\vfill

\begin{center}
{\LARGE\bf  On Consistent Equations for Massive Spin-2 Field Coupled to
Gravity in String Theory}\\

\vfill
%\bigskip

{\sc I.L. Buchbinder${}^a$, V.A. Krykhtin${}^b$ and V.D.
Pershin${}^c$
\footnote{e-mail: \tt ilb@mail.tomsknet.ru,
krykhtin@phys.dfe.tpu.edu.ru, pershin@ic.tsu.ru}
\\

\bigskip

\it ${}^a$Department of Theoretical Physics, Tomsk State Pedagogical
University, \it Tomsk 634041, Russia\\

\bigskip

\it ${}^b$Department of Theoretical and Experimental Physics,
\it Tomsk Polytechnical University,\\
\it Tomsk 634050, Russia \\

\bigskip

\it ${}^c$Department of Theoretical Physics, Tomsk State
University, \it Tomsk 634050, Russia }

\vfill

%\vspace{0.5cm}

\begin{abstract}
We investigate the problem of derivation of consistent equations of
motion for the massive spin 2 field interacting with gravity within
both field theory and string theory. In field theory we derive the
most general classical action with non-minimal couplings in arbitrary
spacetime dimension, find the most general gravitational
background on which this action describes a consistent theory and
generalize the analysis for the coupling with background scalar
dilaton field. We show also that massive spin 2 field allows in
principle consistent description in arbitrary background if one
builds its action in the form of an infinite series in the inverse
mass square. Using sigma-model description of string theory in
background fields we obtain in the lowest order in $\alpha'$ the
explicit form of effective equations of motion for the massive spin 2
field interacting with gravity from the requirement of quantum Weyl
invariance and demonstrate that they coincide with the general form
of consistent equations derived in field theory.
\end{abstract}
\end{center}

\vfill

\newpage

String theory contains an infinite number of massive fields with
various spins interacting with each other and with a finite number of
massless fields and should provide a consistent description of
higher spins interaction. Within ordinary field theory
consistent classical actions for the higher spin fields are known
only on specific curved spacetime manifolds. For example, massive
integer spins described by symmetric tensors of corresponding ranks
were investigated only in spacetimes of constant curvature
\cite{aragone}--\cite{klish2}\footnote{For description of higher spin
massless fields on specific background see e.g.  \cite{vasilev}}.
It means that gravity field in these descriptions should not be
dynamical since it does not feel the presence of higher spins matter
through an energy-momentum tensor.  Hence a consistent classical
action for gravity and a higher massive field is not known and
there are some indications that it does not exist at all.

One of these indications comes from considering a Kaluza-Klein
decomposition of Einstein gravity in $D-$dimensional
spacetime into gravity plus infinite tower of massive second rank
tensor fields in $(D-1)-$dimensional world, masses being
proportional to inverse compactification radius. The resulting four
dimensional theory of spin 2 fields interacting with gravity and with
each other should be consistent as we started from the ordinary
Einstein theory and just consider it on a specific manifold.  But as
was shown in \cite{duff}, it is impossible to reduce this theory
consistently to a finite number of spin 2 fields, i.e. consistency can
be achieved only if the infinite number of higher massive fields are
present in the theory.

From the other hand, there are arguments that in string theory a
general coordinate invariant effective field action reproducing the
correct S-matrix both for massless and massive string states does not
exist too \cite{versus}. The full effective action for all string
fields is not general coordinate invariant and general covariance
arises only as approximate symmetry in effective action for
massless fields once all the massive fields are integrated out.
That effective action for massive fields cannot be covariant follows,
for example, from the fact that terms cubic in massive fields can
contain only flat metric and there is no terms of higher powers
\cite{versus}.

Though influence of massive string modes is negligible at low
energies it is desirable to understand how their consistent
description in string theory arises. In addition to the general
importance of this problem knowledge of effective equations of motion
for massive string modes can be useful in various applications of
string theory, for instance, in  pre-Big Bang cosmology
\cite{maggiore}.

And indeed, as we show in this paper there exists a possibility to
derive from the string theory covariant equations for massive higher
spins fields interacting with background gravity linear in massive
fields. This possibility does not contradict to non-covariance of the
full string field action \cite{versus}. Terms in this action
cubic in massive fields should really be non-covariant but one can
derive in general covariant form terms quadratic in massive fields.
The aim of our paper is to show explicitly how this procedure works
using as an example dynamics of the second rank tensor from the first
massive level of open bosonic string.

A convenient method of deriving effective field equations of motion
from the string theory is provided by the $\sigma-$model approach
\cite{sigma}. Within this approach a string
interacting with background fields is described by a two dimensional
field theory and effective equations of motion arise from the
requirement of quantum Weyl invariance.  Perturbative derivation of
these equations is well suited for massless string modes because the
corresponding two dimensional theory is renormalizable and loop
expansion corresponds to expansion of string effective action in
powers of string length $\sqrt{\alpha'}$.

Inclusion of interaction with massive modes
\cite{massive}--\cite{toder} makes the theory
non-re\-nor\-ma\-li\-zab\-le but this fact does not represent a
problem since in string theory one considers the whole infinite
set of massive fields. Infinite number of counterterms needed for
cancellation of divergences generating by a specific massive field in
classical action leads to renormalization of an infinite number of
massive fields. The only property of the theory crucial for
possibility of derivation of perturbative information is that number
of massive fields giving contributions to renormalization of the
given field should be finite.  As was shown in \cite{bflp} string
theory does fulfill this requirement. To calculate $\beta-$function
for any massive field it is sufficient to find divergences coming
only from a finite number of other massive fields and thus it is
possible to derive perturbatively effective equations of motion for
any background fields in any order in $\alpha'$. To find
non-perturbative contribution to the effective equations of motion
one should use, for example, the method  of exact renormalization
group \cite{ERG}. But this method has a serious disadvantage as it
requires explicit separation of classical action into free and
interaction parts and thus leads to non-covariant equations for
background fields.  From the general point of view this fact does not
contradict general properties of string theory but makes it rather
difficult to establish relations between string fields equations and
ordinary field theory.

In this paper we discuss derivation of covariant equations of
motion for massive string fields interacting with gravity by means of
ordinary perturbative analysis of quantum Weyl invariance condition
in the corresponding $\sigma-$model. Of course, perturbatively we can
obtain equations only linear in massive fields. It was noted long
ago \cite{versus} that one can make such a field redefinition in
the string effective action that terms quadratic in massive fields
(linear terms in equations of motion) acquire dependence on arbitrary
higher powers of massless fields and so may be covariant. In this
paper we receive explicitly these interaction terms in the
lowest in $\alpha'$ approximation.

As a model for our calculations we use bosonic open string theory
interacting with background fields of the massless and the first
massive levels. First massive level contains symmetric second rank
tensor which provides the simplest example of massive higher spin
field interacting with gravity. We begin with analyzing the problem
of consistency for this field interacting with gravity from the point
of view of field theory. Consistency of the theory requires
that number of physical propagating degrees of freedom should be the
same as in the flat space limit. We obtain the most general form of
classical action fulfilling this requirement in arbitrary dimension
and show that in case of actions quadratic in spacetime derivatives
(i.e. linear in curvature) consistency can be achieved only on
gravitational backgrounds with vanishing traceless part of Ricci
tensor. If one includes in the action an infinite number of terms
with all possible powers of curvature then consistency can be
achieved without any restrictions on the background geometry and this
is just the case realized by string theory. In order to obtain
equations of motion for the massive spin 2 field from the
string theory we build effective action for the corresponding two
dimensional theory, perform renormalization of background fields and
composite operators and construct the renormalized operator of energy
momentum tensor trace.

This rather standard scheme was first developed for calculations in
closed string theory in massless background fields
\cite{sigma,composite}
and then was generalized for the open string theory
\cite{open} and for strings in massive fields
\cite{bflp}.

Making perturbative calculations we restrict ourselves to string
world sheets with topology of a disk. The resulting equations of
motion for graviton will not contain dependence on massive fields
from the open string spectrum because these fields interact only with
the boundary of world sheet and can not influence the local
physics in the bulk. For example, in the case of graviton and
massive fields from the open string spectrum one expects that
equations of motion for the graviton would look like ordinary vacuum
Einstein equations without matter. One can obtain contributions from
open string background fields to the right hand side of Einstein
equations for gravity only considering world sheets of higher
genus \cite{loops}.

The organization of the paper is as follows. First we
describe the most general consistent equations of motion for massive
spin 2 field on curved manifolds from the point of view of ordinary
field theory.  Then we describe the string model that
we use for derivations of effective equations of motion of string
massive fields. Requirement of quantum Weyl invariance
leads to effective string fields equations of motion and we show
in the lowest order in $\alpha'$ that these equations are
consistent.

It follows from the analysis of irreducible representations of
4-dimensional Poincare group that massive spin 2 field in flat
spacetime can be described by symmetric transverse and traceless
tensor of the second rank $H_{\mu\nu}$ satisfying mass-shell
condition:
\begin{equation}
\Bigr(\partial^2-m^2\Bigl) H_{\mu\nu}=0 {,}\qquad
\partial^\mu H_{\mu\nu}=0 {,}\qquad
H^\mu{}_\mu=0 {.}
\label{irred}
\end{equation}
In higher dimensional spacetimes Poincare algebras have more than two
Casimir operators and there are several different spins for $D>4$.
Talking about spin 2 massive field in arbitrary dimension we will
mean, as usual, that this field by definition satisfies the same
equations (\ref{irred}) as in $D=4$. After dimensional reduction to
$D=4$ such a field will describe massive spin two representation of
$D=4$ Poincare algebra plus infinite tower of Kaluza-Klein
descendants.

It is well known that all the equations (\ref{irred}) can be derived
from the Fierz-Pauli action:
\begin{eqnarray}
S&=&\int\! d^D x \biggl\{ \frac{1}{4} \partial_\mu H \partial^\mu H
-\frac{1}{4} \partial_\mu H_{\nu\rho} \partial^\mu H^{\nu\rho}
-\frac{1}{2} \partial^\mu H_{\mu\nu} \partial^\nu H
+\frac{1}{2} \partial_\mu H_{\nu\rho} \partial^\rho H^{\nu\mu}
\nonumber\\&&
\qquad\qquad
{} - \frac{m^2}{4} H_{\mu\nu} H^{\mu\nu} + \frac{m^2}{4}  H^2
 \biggr\} {.}
\label{actfield}
\end{eqnarray}
where $H=\eta^{\mu\nu} H_{\mu\nu}$.

Equations of motion following from the action (\ref{actfield})
\begin{eqnarray}
E_{\mu\nu}&=&\partial^2 H_{\mu\nu} - \eta_{\mu\nu} \partial^2 H +
\partial_\mu \partial_\nu H
+ \eta_{\mu\nu} \partial^\alpha \partial^\beta H_{\alpha\beta}
- \partial_\sigma \partial_\mu H^\sigma{}_\nu
- \partial_\sigma \partial_\nu H^\sigma{}_\mu
\nonumber
\\&&\qquad\qquad
{}-m^2 H_{\mu\nu} + m^2 H \eta_{\mu\nu} = 0
\end{eqnarray}
can be used to build $D+1$ expressions without second derivatives of
$H_{\mu\nu}$:
\begin{eqnarray}
&&\partial^\mu E_{\mu\nu} =
m^2 \partial_\nu H - m^2 \partial^\mu H_{\mu\nu} =0
\label{con1}
\\
&&\frac{m^2}{D-2} \eta^{\mu\nu} E_{\mu\nu}
+ \partial^\mu \partial^\nu E_{\mu\nu} =
  H  m^4 \frac{D-1}{D-2} =0
\label{con2}
\end{eqnarray}
These expressions represent constraints on the initial values for the
field $H_{\mu\nu}$ and its first derivatives.
Thus the theory contains the same
%$\frac{D(D+1)}{2}-D-1=\frac{(D+1)(D-2)}{2}$
local dynamical degrees
of freedom as the system (\ref{irred}) and describes traceless and
transverse symmetric tensor field of the second rank.

Now if we want to construct a theory of massive spin 2 field on a
curved manifold we should provide the same number of propagating
degrees of freedom as in the flat case. It means that one should be
able to build exactly $D+1$ constraints from the equations of motion
$E_{\mu\nu}$ and in the flat spacetime limit these constraints should
reduce to (\ref{con1},\ref{con2}).

Generalizing (\ref{actfield}) to curved spacetime we should substitute
all derivatives for the covariant ones and also we can add
non-minimal terms containing curvature tensor. As a result,
the most general action for massive spin 2 field in curved spacetime
quadratic in derivatives and consistent with the flat limit
should have the form \cite{aragone}
\begin{eqnarray}&&
S=\int d^D x\sqrt{-G} \biggl\{ \frac{1}{4} \nabla_\mu H \nabla^\mu H
-\frac{1}{4} \nabla_\mu H_{\nu\rho} \nabla^\mu H^{\nu\rho}
-\frac{1}{2} \nabla^\mu H_{\mu\nu} \nabla^\nu H
+\frac{1}{2} \nabla_\mu H_{\nu\rho} \nabla^\rho H^{\nu\mu}
\nonumber
\\&&
\qquad
{}+\frac{a_1}{2} R H_{\alpha\beta} H^{\alpha\beta}
+\frac{a_2}{2} R H^2
+\frac{a_3}{2} R^{\mu\alpha\nu\beta} H_{\mu\nu} H_{\alpha\beta}
+\frac{a_4}{2} R^{\alpha\beta} H_{\alpha\sigma} H_\beta{}^\sigma
+\frac{a_5}{2} R^{\alpha\beta} H_{\alpha\beta} H
\nonumber
\\&&
\qquad
{}- \frac{m^2}{4} H_{\mu\nu} H^{\mu\nu} + \frac{m^2}{4} H^2
 \biggr\}
\label{genact}
\end{eqnarray}
where $a_1, \ldots a_5$ are so far arbitrary coefficients,
$R^\mu{}_{\nu\lambda\kappa}=\partial_\lambda \Gamma^\mu_{\nu\kappa}
-\ldots$; $R_{\mu\nu}=R^\lambda_{\mu\lambda\nu}$.

Equations of motion
\begin{eqnarray}
E_{\mu\nu}&=&\nabla^2 H_{\mu\nu} - G_{\mu\nu} \nabla^2 H +
\nabla_\mu \nabla_\nu H
+ G_{\mu\nu} \nabla^\alpha \nabla^\beta H_{\alpha\beta}
- \nabla_\sigma \nabla_\mu H^\sigma{}_\nu
- \nabla_\sigma \nabla_\nu H^\sigma{}_\mu
\nonumber
\\&&
{}+2a_1 R H_{\mu\nu}
+2a_2 G_{\mu\nu} R H
+2a_3 R_\mu{}^\alpha{}_\nu{}^\beta H_{\alpha\beta}
+a_4 R_\mu{}^\alpha H_{\alpha\nu}
+a_4 R_\nu{}^\alpha H_{\alpha\mu}
\nonumber
\\&&
{}+a_5 R_{\mu\nu} H
+a_5 G_{\mu\nu} R^{\alpha\beta} H_{\alpha\beta}
-m^2 H_{\mu\nu} + m^2 H G_{\mu\nu} = 0
\label{lagreq}
\end{eqnarray}
should contain one vector constraint and one scalar constraint
generalizing (\ref{con1},\ref{con2}) for the case of curved
background.

There are no problems with generalization of the vector constraint
(\ref{con1}). It does not contain second derivatives of $H_{\mu\nu}$
for any gravitational background and for any values of coefficients
$a_1, \ldots a_5$:
\begin{eqnarray}
\nabla^\mu E_{\mu\nu} &=&
2a_1 R \nabla^\mu H_{\mu\nu}
+2a_2 R \nabla_\nu H
+2a_3 R^{\mu\alpha}{}_\nu{}^\beta \nabla_\mu H_{\alpha\beta}
+a_4 R^{\mu\alpha} \nabla_\mu H_{\alpha\nu}
\nonumber\\&&{}
+(a_4-2) R^\alpha{}_\nu \nabla^\mu H_{\alpha\mu}
+a_5 R^{\alpha\mu} \nabla_\nu H_{\alpha\mu}
+(a_5+1) R^\alpha{}_\nu \nabla_\alpha H
\nonumber\\&&{}
{}- m^2 \nabla^\mu H_{\mu\nu} + m^2 \nabla_\nu H +
\ldots =0
\end{eqnarray}
where dots stand for the terms not containing derivatives
of the field $H_{\mu\nu}$.

As for the curved space generalization of the scalar constraint
(\ref{con2}), it can have additional terms proportional to curvature
and does contain second derivatives:
\begin{eqnarray}
&& \frac{m^2}{D-2} G^{\mu\nu} E_{\mu\nu}
+ \nabla^\mu \nabla^\nu E_{\mu\nu}
+ b_1 R G^{\mu\nu} E_{\mu\nu}
+ b_2 R^{\mu\nu} E_{\mu\nu} =
\nonumber\\&&{}
= \Bigl(2a_1+b_1(D-2)+b_2 -\frac{2a_3}{D(D-1)}+\frac{2a_4-2b_2-2}{D}\Bigr)
          R \nabla^\alpha \nabla^\beta H_{\alpha\beta}
\nonumber\\&&\qquad{}
+ \Bigl(2a_2-b_1(D-2)-b_2 +\frac{2a_3}{D(D-1)}
          + \frac{2a_5+2b_2+1}{D} \Bigr) R \nabla^2 H
\nonumber\\&&\qquad{}
+ 2a_3 C^{\mu\alpha\nu\beta} \nabla_\mu \nabla_\nu H_{\alpha\beta}
+ \Bigl(2a_4-2b_2-2 - \frac{4a_3}{D-2} \Bigr) \tilde{R}^{\alpha\beta}
        \nabla_\alpha \nabla^\mu H_{\mu\beta}
\nonumber\\&&\qquad{}
+ \Bigl(a_5+b_2 +\frac{2a_3}{D-2} \Bigr) \tilde{R}^{\alpha\beta}
           \nabla^2 H_{\alpha\beta}
+ \Bigl(a_5+b_2+1+ \frac{2a_3}{D-2} \Bigr) \tilde{R}^{\alpha\beta}
\nabla_\alpha \nabla_\beta H + \ldots =0
\label{cancel}
\end{eqnarray}
where dots stand for the terms without second derivatives of $H_{\mu\nu}$
and we decomposed the curvature tensor
$R_{\mu\nu\alpha\beta}$ into Weyl tensor $C_{\mu\nu\alpha\beta}$,
traceless part of the Ricci tensor
$\tilde{R}_{\mu\nu}=R_{\mu\nu}-\frac{1}{D}G_{\mu\nu}R$
and scalar curvature.
It is impossible to cancel all second derivatives in (\ref{cancel})
by fixing the coefficients $a_i$ since at least some combination of
the last two terms with $\tilde{R}^{\alpha\beta}$ will always remain.
So in order to achieve consistency with the flat space limit one has
to impose restriction on the gravitational background
\begin{equation}
R_{\mu\nu}=\frac{1}{D}G_{\mu\nu}R
\label{restr}
\end{equation}
thus cancelling bad terms with second derivatives of $H_{\mu\nu}$ in
(\ref{cancel}). This restriction is no very strong and allows one to
consider a wide class of curved backgrounds. Note, that in all
previous works only the constant curvature D=4 spacetimes were
considered. The condition (\ref{restr}) means that only traceless
part of Ricci tensor should vanish while scalar curvature and Weyl
tensor can be arbitrary.

In fact, this condition can be yet more weakened because
in order to achieve consistency one should cancel in (\ref{cancel})
only second {\em time} derivatives. For example, if we are dealing
with a stationary spacetime possessing a timelike Killing vector
$k^\mu$, $k^2<0$ it is sufficient to demand:
$$ k^\mu k^\nu R_{\mu\nu} = \frac{1}{D} k^2 R $$
which means that only projection of the traceless part of
$R_{\mu\nu}$ on orbits of $k^\mu$ should vanish.

Other terms with second derivatives in (\ref{cancel}) can be
cancelled by choosing appropriate coefficients. As a result, one
arrives at one-parameter family of consistent theories for the
massive spin-2 fields on the spacetime fulfilling (\ref{restr}):
\begin{eqnarray}&&
S=\int d^D x\sqrt{-G} \biggl\{ \frac{1}{4} \nabla_\mu H \nabla^\mu H
-\frac{1}{4} \nabla_\mu H_{\nu\rho} \nabla^\mu H^{\nu\rho}
-\frac{1}{2} \nabla^\mu H_{\mu\nu} \nabla^\nu H
+\frac{1}{2} \nabla_\mu H_{\nu\rho} \nabla^\rho H^{\nu\mu}
\nonumber
\\&&
\qquad\qquad
+\frac{\xi}{2D} R H_{\mu\nu} H^{\mu\nu} +\frac{1-2\xi}{4D} R H^2
- \frac{m^2}{4} H_{\mu\nu} H^{\mu\nu} + \frac{m^2}{4} H^2
 \biggr\} {.}
\label{curvact}
\end{eqnarray}
$\xi$ is the only dimensionless coupling constant
responsible for non-minimality of interaction with curved background.
This is the most general form of the action for the massive spin 2
field in curved spacetime of arbitrary dimension leading to
consistent equations of motion.  Previous works
\cite{aragone}-\cite{cpd} treated only cases with various
particular values of $\xi$ and only for $D=4$.

Equations of motion and constraints have the following form:
\begin{eqnarray}&&
E_{\mu\nu}=\nabla^2 H_{\mu\nu} - G_{\mu\nu} \nabla^2 H +
\nabla_\mu \nabla_\nu H
+ G_{\mu\nu} \nabla^\alpha \nabla^\beta H_{\alpha\beta}
- \nabla_\sigma \nabla_\mu H^\sigma{}_\nu
- \nabla_\sigma \nabla_\nu H^\sigma{}_\mu
\nonumber
\\&&\qquad\qquad
+ \frac{2\xi}{D} R H_{\mu\nu} + \frac{1-2\xi}{2D} RH G_{\mu\nu}
-m^2 H_{\mu\nu} + m^2 H G_{\mu\nu} = 0
\label{curveq}
\\&&
\nabla^\mu E_{\mu\nu} =
\frac{2(1-\xi)}{D} \biggl( R\nabla_\nu H + H
\nabla_\nu R + R \nabla^\mu H_{\mu\nu} + H_{\mu\nu} \nabla^\mu R
\biggr)
\nonumber\\&&\qquad\qquad
- m^2 \nabla^\mu H_{\mu\nu} + m^2 \nabla_\nu H =0
\label{curvcon1}
\\&&
\frac{m^2}{D-2} G^{\mu\nu} E_{\mu\nu}
+ \nabla^\mu \nabla^\nu E_{\mu\nu}
+ \frac{2(1-\xi)}{D(D-2)} R G^{\mu\nu} E_{\mu\nu} =
\nonumber
\\&&\qquad\qquad{}
=  \frac{2(1-\xi)}{D} \biggl(
2\nabla^\mu R \nabla_\mu H + H \nabla^2 R
- 2 \nabla^\mu R \nabla^\nu H_{\mu\nu} -
H_{\mu\nu} \nabla^\mu \nabla^\nu R \biggr)
\nonumber\\&&\qquad\qquad{}
+ H \biggl( m^4 \frac{D-1}{D-2}
+m^2 R \frac{3D-2+2\xi(1-D)}{D(D-2)} +
R^2 \frac{2(1-\xi)(D+2\xi(1-D))}{D^2(D-2)}
\biggr)=0
\label{curvcon2}
\end{eqnarray}
Note that consistency of the theory (in the sense that it has correct
flat limit) cannot be achieved with minimal coupling to gravity. For
any value of $\xi$ there are non-minimal terms in (\ref{curvact}).

The constraints (\ref{curvcon1},\ref{curvcon2}) have the simplest
form when $\xi=1$. The equations of motion in this case reduce to:
\begin{eqnarray}
&&\nabla^2 H_{\mu\nu} + 2 R^\alpha{}_\mu{}^\beta{}_\nu
H_{\alpha\beta} - m^2 H_{\mu\nu} = 0 {,}
\nonumber\\&&
H^\mu{}_\mu=0{,} \qquad \nabla^\mu H_{\mu\nu} = 0 {,}
\label{simple}
\end{eqnarray}
The theory is consistent for any other values of $\xi$, only explicit
form of constraints and dynamical equations of motion is more
complicated than that of (\ref{simple}).

One can generalize the above analysis for the case of spin 2 massive
field interacting not only with background gravity but also with
the scalar dilaton field. This set of fields arises naturally in
string theory which contains  dilaton field $\phi(x)$ as one of its
massless excitations.

Writing a general action similar to (\ref{genact}) for this system
one should take into account all possible new terms
with derivatives of $\phi(x)$ and also containing arbitrary
factors $f(\phi)$ without derivatives of the dilaton fields. For
example, string effective action can contain in various terms the
factors $e^{k\phi}$, $k=const$. We will consider here the class of
actions for the field $H_{\mu\nu}$ where all these factors can be
absorbed to the metric $G_{\mu\nu}$ by a conformal rescaling.

Repeating in this case the above analysis made for spin 2 field
coupled to gravity and demanding cancellation of the second
derivatives in the constraints we found that the most general action
in presence of dilaton can depend on 4 extra parameters  of
non-minimal coupling $\zeta$:
\begin{eqnarray*}
&& S=\int d^D x\sqrt{-G}
\biggl\{ \frac{1}{4} \nabla_\mu H \nabla^\mu H -\frac{1}{4}
\nabla_\mu H_{\nu\rho} \nabla^\mu H^{\nu\rho} -\frac{1}{2} \nabla^\mu
H_{\mu\nu} \nabla^\nu H +\frac{1}{2} \nabla_\mu H_{\nu\rho}
\nabla^\rho H^{\nu\mu} \\&& \qquad{} +\frac{\xi}{2D} R H_{\mu\nu}
H^{\mu\nu} +\frac{1-2\xi}{4D} R H^2
+\frac{\zeta_1}{2}\nabla^\alpha\phi\nabla_\alpha H_{\mu\nu}H^{\mu\nu}
-\frac{\zeta_1}{2}\nabla^\alpha\phi\nabla_\alpha H H \\&& \qquad{}
+\frac{\zeta_2}{2}\nabla^\alpha \phi \nabla_\mu H_{\alpha\nu} H^{\mu\nu}
-\frac{\zeta_2}{2}\nabla^\alpha \phi \nabla^\mu H_{\alpha\mu} H
-\frac{\zeta_2}{2}\nabla^\mu \phi \nabla^\nu H H_{\mu\nu}
+\frac{\zeta_2}{2}\nabla^\alpha \phi \nabla^\mu H_{\mu}{}^\beta H_{\alpha\beta}
\\&&
\qquad{}
+\frac{\zeta_3}{2}\nabla^\mu \phi \nabla^\nu \phi H_{\mu\alpha} H_\nu{}^\alpha
-\frac{\zeta_3}{2}\nabla^\mu \phi \nabla^\nu \phi H_{\mu\nu} H
+\frac{\zeta_4}{2} (\nabla \phi)^2 H_{\mu\nu} H^{\mu\nu}
-\frac{\zeta_4}{2} (\nabla \phi)^2 H^2
\\&&
\qquad{}
- \frac{m^2}{4} H_{\mu\nu} H^{\mu\nu} + \frac{m^2}{4} H^2
 \biggr\} {.}
\label{dilact}
\end{eqnarray*}
The restriction on the gravitational background (\ref{restr})  remains
the same as in the case without dilaton and no restrictions on the
dilaton arise.  There exist $D+1$ constraints built out of equations of
motion $E_{\mu\nu}=0$: $D$ constraints
$$\nabla^\mu E_{\mu\nu}=0$$
and one scalar constraint
$$
\frac{m^2}{D-2} G^{\mu\nu} E_{\mu\nu}
+ \nabla^\mu \nabla^\nu E_{\mu\nu}
+ \frac{2(1-\xi)}{D(D-2)} R G^{\mu\nu} E_{\mu\nu}
$$
$$
-\zeta_2 \nabla^\mu \nabla^\nu \phi E_{\mu\nu}
+\zeta_3 \nabla^\mu \phi \nabla^\nu \phi E_{\mu\nu}
- \frac{\zeta_3+2\zeta_4}{D-2} (\nabla\phi)^2 G^{\mu\nu} E_{\mu\nu}
+ \frac{\zeta_1+\zeta_2}{D-2} \nabla^2 \phi G^{\mu\nu} E_{\mu\nu} =0
$$
The scalar dilaton here is  considered as background field on the same
footing as $G_{\mu\nu}$. In the rest of the paper we will consider the
case when the dilaton is absent and take into account  only pure
gravitational background.

Of course, one would like to overcome the restriction (\ref{restr})
on the background gravitational field and to construct consistent
theory of the massive spin 2 in arbitrary curved spacetime. Such a
possibility really exists if one considers a lagrangian which contains
an infinite number of terms.  It is possible only for massive fields
because the mass of the field $H_{\mu\nu}$ is the only dimensionful
parameter which can be used to construct a lagrangian with terms of
arbitrary orders in curvature multiplied by the corresponding powers of
$1/m^2$:
\begin{eqnarray}&&
S_H =\int d^D x\sqrt{-G} \biggl\{ \frac{1}{4} \nabla_\mu H \nabla^\mu H
-\frac{1}{4} \nabla_\mu H_{\nu\rho} \nabla^\mu H^{\nu\rho}
-\frac{1}{2} \nabla^\mu H_{\mu\nu} \nabla^\nu H
+\frac{1}{2} \nabla_\mu H_{\nu\rho} \nabla^\rho H^{\nu\mu}
\nonumber
\\&&
\qquad
{}+\frac{a_1}{2} R H_{\alpha\beta} H^{\alpha\beta}
+\frac{a_2}{2} R H^2
+\frac{a_3}{2} R^{\mu\alpha\nu\beta} H_{\mu\nu} H_{\alpha\beta}
+\frac{a_4}{2} R^{\alpha\beta} H_{\alpha\sigma} H_\beta{}^\sigma
+\frac{a_5}{2} R^{\alpha\beta} H_{\alpha\beta} H
\nonumber\\&&\qquad{}
+\frac{1}{m^2} ( R \nabla H \nabla H + R H \nabla\nabla H + R R H H)
+ O\Bigl(\frac{1}{m^4}\Bigr)
- \frac{m^2}{4} H_{\mu\nu} H^{\mu\nu} + \frac{m^2}{4} H^2
 \biggr\}
\label{higher}
\end{eqnarray}
Actions of this kind are expected to arise naturally in string
theory where the role of mass parameter is played by string
tension $m^2=1/\alpha'$ and perturbation theory in $\alpha'$
gives effective actions of the form (\ref{higher})\footnote{In fact,
string theory should lead to even more general effective actions than
(\ref{higher}) since in the higher $\alpha'$ corrections higher
derivatives of the field $H_{\mu\nu}$ should appear.}.

Possibility of constructing consistent equations for massive higher
spin fields as series in curvature was recently studied in
\cite{klish2} where for spin 2 field these equations were derived
in particular case of symmetrical Einstein spaces in linear in
curvature order. In principle, as we argue below consistency can be
achieved without imposing any restrictions on gravitational
background, at least in the lowest order in $1/m^2$.

Equations of motion are:
\begin{eqnarray}
E_{\mu\nu}&=&\nabla^2 H_{\mu\nu} - G_{\mu\nu} \nabla^2 H +
\nabla_\mu \nabla_\nu H
+ G_{\mu\nu} \nabla^\alpha \nabla^\beta H_{\alpha\beta}
- \nabla_\sigma \nabla_\mu H^\sigma{}_\nu
- \nabla_\sigma \nabla_\nu H^\sigma{}_\mu
\nonumber
\\&&
{}+2a_1 R H_{\mu\nu}
+2a_2 G_{\mu\nu} R H
+2a_3 R_\mu{}^\alpha{}_\nu{}^\beta H_{\alpha\beta}
+a_4 R_\mu{}^\alpha H_{\alpha\nu}
+a_4 R_\nu{}^\alpha H_{\alpha\mu}
\nonumber\\&&{}
+a_5 R_{\mu\nu} H
+a_5 G_{\mu\nu} R^{\alpha\beta} H_{\alpha\beta}
+\frac{1}{m^2} (R\nabla\nabla H+ \nabla R \nabla H + \nabla\nabla R H
+ R R H)
\nonumber\\&&{}
+ O \left(\frac{1}{m^4}\right)
-m^2 H_{\mu\nu} + m^2 H G_{\mu\nu} = 0
\label{eqhigh}
\end{eqnarray}
Now we solve perturbatively the constraints with respect to the trace
and longitudinal part of $H_{\mu\nu}$. Of course, the combinations of
equations of motion $E_{\mu\nu}$ forming the constraints can change
in higher orders but this is irrelevant for the lowest order
conditions. The vector constraint arise from:
\begin{eqnarray}
&&\nabla^\mu E_{\mu\nu} + \frac{1}{m^2} (R\nabla E+ \nabla R E) +
O \left(\frac{1}{m^4}\right) =
-m^2 \nabla^\mu H_{\mu\nu} + m^2 \nabla_\nu H + O(1) =0
\qquad\Rightarrow
\nonumber\\&&
\qquad\qquad\qquad\qquad\Rightarrow\qquad
\nabla^\mu H_{\mu\nu} - \nabla_\nu H  = O\left(\frac{1}{m^2}\right)
\end{eqnarray}

Since the trace enters the equations of motion in the term $m^2 H$ we
will need expansion of the scalar constraint up to the next to
leading terms:
\begin{eqnarray}
&& \frac{m^2}{D-2} G^{\mu\nu} E_{\mu\nu}
+ \nabla^\mu \nabla^\nu E_{\mu\nu}
+ b_1 R G^{\mu\nu} E_{\mu\nu}
+ b_2 R^{\mu\nu} E_{\mu\nu}+ O \left(\frac{1}{m^2}\right) =
\nonumber\\&&\qquad{}
= m^4 \frac{D-1}{D-2} H
+ m^2 RH \Bigl[ \frac{2a_1+2Da_2+a_5}{D-2} + (D-1)b_1 + b_2 \Bigr]
\nonumber\\&&\qquad\qquad{}
+ m^2 R^{\alpha\beta} H_{\alpha\beta}
\Bigl[\frac{2a_3+2a_4+Da_5}{D-2} - b_2 \Bigr] + O(1)
= 0 {,}
\label{x}
\end{eqnarray}
We see that in the lowest orders the expression (\ref{x}) does not
contain second derivatives for any values of the coefficients $a_i$,
$b_i$ and for arbitrary background.  Solving it with respect to the
trace we get
\begin{equation}
H= \frac{1}{m^2} R^{\alpha\beta} H_{\alpha\beta}
\frac{(D-2)b_2 - 2a_3-2a_4-Da_5}{D-1}
+ O \left(\frac{1}{m^4}\right)
\label{trace}
\end{equation}

Substituting (\ref{trace}) back to the equations of motion we get in
the lowest order the system of equations for $H_{\mu\nu}$ equivalent
to the original equations (\ref{eqhigh}):
\begin{eqnarray}
&&\nabla^2 H_{\mu\nu}
+ 2a_1 R H_{\mu\nu}
+ (2a_3+2) R^\alpha{}_\mu{}^\beta{}_\nu H_{\alpha\beta}
+ 2 (a_4-1) R^\sigma{}_{(\mu} H_{\nu)\sigma}
\nonumber\\&&\qquad{}
+  G_{\mu\nu} R^{\alpha\beta} H_{\alpha\beta}
 \frac{(D-2)b_2-2a_3-2a_4-a_5}{D-1}
- m^2 H_{\mu\nu} + O \left(\frac{1}{m^2}\right) = 0 {,}
\nonumber\\&&{}
\nabla^\mu  H_{\mu\nu} + O \left(\frac{1}{m^2}\right) =0 {,}
\qquad
H_\mu{}^\mu + O \left(\frac{1}{m^2}\right) = 0 {.}
\label{finhigh}
\end{eqnarray}
Mass shell condition in (\ref{finhigh}) contains in this order four
non-minimal terms with curvature tensor and coefficients at all of
them are arbitrary.

This arbitrariness is related to the possibility of making field
redefinitions in the action (\ref{higher}). It is well known
\cite{tsPLB} that in string theory such possibility allows to change
effective action for massless background fields in higher orders. In
case of massive fields action contains the mass term which is of
order $1/\alpha'$ and this leads to possibility of changing some
coefficient in the effective action already in the lowest order.
Namely, since (\ref{higher}) contains all higher powers of $1/m^2$
one can redefine the field $H_{\mu\nu}$ through an infinite series
as:
\begin{equation}
H_{\mu\nu} \rightarrow  H_{\mu\nu} + \frac{1}{m^2}
\bigl(
 \lambda_1 R_\mu{}^\alpha{}_\nu{}^\beta H_{\alpha\beta}
+ \lambda_2 R^\alpha{}_{(\mu} H_{\nu)\alpha}
+ \lambda_3 R H_{\mu\nu}
+ \lambda_4 G_{\mu\nu} R^{\alpha\beta} H_{\alpha\beta}
\bigr)
+ O \left(\frac{1}{m^4}\right)
\label{redef}
\end{equation}
Using such redefinitions one can assign any values to the
coefficients $a_1$, $a_3$, $a_4$, $b_2$ in (\ref{finhigh}) and so
specific values of these coefficients do not play any role in
achieving consistency of the theory (\ref{higher}). Requirement of
consistency will restrict parameters of the theory only in higher
orders in $1/m^2$.

We would like to stress once more that
the theory (\ref{higher}) admits any gravitational background and so
no inconsistencies arise if one treats gravity as dynamical field
satisfying Einstein equations with the energy - momentum tensor for
the field $H_{\mu\nu}$. This system is consistent in the sense that
it has correct number of degrees of freedom.  One can also study
additional requirements the theory should fulfill, e.g. causality
\cite{aragone} or tree level unitarity of graviton - massive spin 2
field interaction \cite{cpd}.  These requirements can lead to some
additional restrictions on the parameters of the theory.

Now we consider sigma-model description of an
open string interacting with two background fields -- massless
graviton $G_{\mu\nu}$ and second rank symmetric tensor field
$H_{\mu\nu}$ from the first massive level of the open string
spectrum. We are going to show that effective equations of motion for
these fields are of the form (\ref{finhigh}) and explicitly calculate
all the coefficients in these equations in the lowest order in
$\alpha'$.

Classical action has the form
\begin{equation}
S=S_0+S_I=
\frac{1}{4\pi\alpha'}\int_M \!\!d^2z\sqrt{g}
      g^{ab}\partial_ax^\mu\partial_bx^\nu G_{\mu\nu}
+\frac{1}{2\pi\alpha'\mu}\int_{\partial M} dt \; e
      H_{\mu\nu}\dot{x}^\mu\dot{x}^\nu
\label{actstring}
\end{equation}
Here $\mu,\nu=0,\ldots,D-1$; $a,b=0,1$ and we introduced the notation
$\dot{x}^\mu=\frac{dx^\mu}{edt}$. The first term $S_0$ is an integral
over two-dimensional string world sheet $M$ with metric $g_{ab}$ and
the second $S_I$ represents a one-dimensional integral over its
boundary with einbein $e$. We work in euclidian signature and
restrict ourselves to flat world sheets with straight boundaries. It
means that both two-dimensional scalar curvature and extrinsic
curvature of the world sheet boundary vanish and we can always choose
such coordinates that $g_{ab}=\delta_{ab}$, $e=1$.

Theory has two dimensionful parameters. $\alpha'$ is fundamental
string length squared, $D$-dimensional coordinates $x^\mu$ have
dimension $\sqrt{\alpha'}$. Another parameter $\mu$ carries dimension
of inverse length in two-dimensional field theory (\ref{actstring})
and plays the role of renormalization scale. It is introduced in
(\ref{actstring}) to make the background field $H_{\mu\nu}$
dimensionless. In fact, power of $\mu$ is responsible for the number
of massive level to which a background field belongs because one
expects that open string interacts with a field from $n$-th massive
level by means of the term
$$
\mu^{-n} (\alpha')^{-\frac{n+1}{2}}
\int_{\partial M}  dt\; e \dot{x}{}^{\mu_1} \ldots
\dot{x}{}^{\mu_{n+1}} H_{\mu_1 \ldots \mu_{n+1}} (x)
$$

The action (\ref{actstring}) is non-renormalizable from the point of
view of two-dimensional quantum field theory. Inclusion of
interaction with any massive background produces in each loop an
infinite number of divergencies and requires an infinite number of
different massive fields in the action. But massive modes from the
$n-$th massive level give vertices proportional to $\mu^{-n}$ and so
they cannot contribute to renormalization of fields from lower
levels. Of course, this argument supposes that we treat the theory
perturbatively defining propagator for $X^\mu$ only by the graviton
term in (\ref{actstring}).

Varying (\ref{actstring}) one gets classical equations of motion with
boundary conditions:
\begin{eqnarray}
&& g^{ab} D_a \partial_b x^\alpha \equiv
g^{ab}(\partial_a \partial_b x^\alpha + \Gamma^\alpha_{\mu\nu}(G)
\partial_a x^\mu \partial_b x^\nu) = 0,
\nonumber\\&&
\left. G_{\mu\nu} \partial_n x^\mu \right|_{\partial M} -
\frac{2}{\mu} {\cal D}^2_t x^\mu H_{\mu\nu} +\frac{1}{\mu}
{\dot x}^\mu {\dot x}^\lambda (\nabla_\nu H_{\mu\lambda}
-\nabla_\mu H_{\nu\lambda} - \nabla_\lambda H_{\mu\nu}) = 0
\label{class}
\end{eqnarray}
where  $\partial_n=n^a \partial_a$, $n^a$ -- unit inward normal
vector to the world sheet boundary and ${\cal D}^2_t x^\mu = {\ddot
x}^\mu+\Gamma^\mu_{\nu\lambda}(G) {\dot x}^\nu {\dot x}^\lambda$.

Divergent part of the one loop effective action has the form
\begin{eqnarray}
&&\Gamma^{(1)}_{div}=\frac{\mu^{-\varepsilon}}{4\pi\varepsilon}
    \int_{M} d^{2+\varepsilon} z\sqrt{g}
    g^{ab}\partial_ax^\mu\partial_bx^\nu R_{\mu\nu}
\nonumber\\&&\qquad{}
    -\frac{\mu^{-\varepsilon-1}}{2\pi\varepsilon}\int_{\partial M}
    dt e(t) \dot{x}^\mu\dot{x}^\nu
   \left( \nabla^2 H_{\mu\nu} - 2R_\mu{}^\alpha H_{\alpha\nu}
+ R_\mu{}^\alpha{}_\nu{}^\beta H_{\alpha\beta} \right)
+ O(\mu^{-2})
\end{eqnarray}
where the terms $O(\mu^{-2})$ give contributions to
renormalization of only the second and higher massive levels.
Hence one-loop renormalization of the background fields looks like:
\begin{eqnarray}
\stackrel{\circ}{G}_{\mu\nu}&=&\mu^\varepsilon
  G_{\mu\nu}-\frac{\alpha'\mu^\varepsilon}{\varepsilon}R_{\mu\nu}
\nonumber\\
\stackrel{\circ}{H}_{\mu\nu}&=&\mu^\varepsilon H_{\mu\nu}
  +\frac{\alpha'\mu^\varepsilon}{\varepsilon}
  \left( \nabla^2 H_{\mu\nu} -2 R^\sigma{}_{(\mu} H_{\nu)\sigma}
  + R_\mu{}^\alpha{}_\nu{}^\beta H_{\alpha\beta} \right)
\label{1loop}
\end{eqnarray}
with circles denoting bare values of the fields. We would like to
stress once more that higher massive levels do not influence the
renormalization of any given field from the lower massive levels and
so the result (\ref{1loop}) represents the full answer for
perturbative one-loop renormalization of $G_{\mu\nu}$ and
$H_{\mu\nu}$.

Now to impose the condition of Weyl invariance of the theory at the
quantum level we calculate the trace of energy momentum tensor
in $d=2+\varepsilon$ dimension:
\begin{equation}
  T(z) = g_{ab}(z) \frac{\delta S}{\delta g_{ab}(z)} =
\frac{\varepsilon\mu^{-\varepsilon}}{8\pi\alpha'}
g^{ab}(z)\partial_ax^\mu\partial_bx^\nu G_{\mu\nu}
-\frac{\mu^{-1-\varepsilon}}{4\pi\alpha'} H_{\mu\nu} \dot{x}^\mu
\dot{x}^\nu \delta_{\partial M}(z)
\end{equation}
and perform one-loop renormalization of the composite operators:
\begin{equation}
\bigl( \dot{x}^\mu \dot{x}^\nu \stackrel{\circ}{H}_{\mu\nu} \bigr)_0 =
\mu^{-\varepsilon}
\bigl[ \dot x^\mu \dot x^\nu  H_{\mu\nu} \bigr]
\end{equation}
\begin{eqnarray}&&
(g^{ab}\partial_a x^\mu \partial_b x^\nu
   \stackrel{\circ}{G}_{\mu\nu})_0 =\mu^\varepsilon
   \Bigl[g^{ab}\partial_a x^\mu \partial_b x^\nu (G_{\mu\nu}
   -\frac{\alpha'}{\varepsilon} R_{\mu\nu})\Bigr]
\\&&\qquad\qquad{}
+\frac{\alpha'\mu^{-1+\varepsilon}}{\varepsilon}
    \left[H_{\alpha}{}^\alpha\delta_{\partial M}''(z)
   + {\cal D}^2_tx^\mu (\nabla_\mu H_\alpha{}^\alpha
   - 4 \nabla^\alpha H_{\alpha\mu}) \delta_{\partial M}(z)
\right.
\nonumber\\&&\qquad\qquad{}
\left.
   + {\dot x}^\mu {\dot x}^\nu (
    \nabla_\mu \nabla_\nu H_\alpha{}^\alpha
   - 4 \nabla^\alpha \nabla_{(\mu} H_{\nu)\alpha}
   + 2 \nabla^2 H_{\mu\nu} - 2 R_\mu{}^\alpha{}_\nu{}^\beta
     H_{\alpha\beta} ) \delta_{\partial M}(z) \right]
\nonumber
\end{eqnarray}
Here delta-function of the boundary $\delta_{\partial M}(z)$ is
defined as
\begin{eqnarray}
&& \int_M\delta_{\partial M}(z)
V(z)\sqrt{g(z)} d^2z =\int_{\partial M}V|_{z\in\partial M} e(t) dt
\end{eqnarray}

The renormalized operator of the energy momentum tensor
trace is:
\begin{eqnarray}
8\pi[T]&=&{}
- \Bigl[ g^{ab} \partial_a x^\mu \partial_b x^\nu
E_{\mu\nu}^{(0)}(x)\Bigr] + \frac{2}{\mu} \delta_{\partial M}(z)
   \Bigl[ {\dot x}^\mu {\dot x}^\nu E^{(1)}_{\mu\nu}(x) \Bigr]
\nonumber\\&&{}
+ \frac{1}{\mu} \delta_{\partial M}(z)
   \Bigl[ {\cal D}^2_t x^\mu E^{(2)}_\mu(x) \Bigr]
+ \frac{1}{\mu} \delta''_{\partial M}(z) \Bigl[ E^{(3)}(x) \Bigr]
\end{eqnarray}
where
\begin{eqnarray}
E^{(0)}_{\mu\nu} (x)&=& R_{\mu\nu} + O(\alpha')
\nonumber\\
E^{(1)}_{\mu\nu}(x)&=&
\nabla^2 H_{\mu\nu} - \nabla^\alpha \nabla_\mu H_{\alpha\nu}
- \nabla^\alpha \nabla_\nu H_{\alpha\mu}
%\nonumber\\&&{}
- R_\mu{}^\alpha{}_\nu{}^\beta H_{\alpha\beta}
+ \frac{1}{2} \nabla_\mu \nabla_\nu H_\alpha{}^\alpha
- \frac{1}{\alpha'} H_{\mu\nu}  + O(\alpha')
\nonumber\\
E^{(2)}_\mu(x)&=& \nabla_\mu H_\alpha{}^\alpha
- 4 \nabla^\alpha H_{\alpha\mu} + O(\alpha')
\nonumber\\
E^{(3)}(x)&=&H_\alpha{}^\alpha + O(\alpha')
\label{E}
\end{eqnarray}
Terms of order $O(\alpha')$ arise from the higher loops contributions.

The requirement of quantum Weyl invariance tells that all $E(x)$
should vanish and so they are interpreted as effective
equations of motion for background fields. They contain vacuum
Einstein equation for graviton (in the lowest order in $\alpha'$),
curved spacetime generalization of the mass shell condition for the
field $H_{\mu\nu}$ with the mass $m^2=(\alpha')^{-1}$ and $D+1$
additional constraints on the values of this fields and its first
derivatives. In fact, Einstein equations should not be vacuum ones
but contain dependence on the field $H_{\mu\nu}$ through its energy -
momentum tensor $T^H_{\mu\nu}$. Our calculations could not produce this
dependence because it is expected to arise only if one
takes into account string world sheets with non-trivial topology and
renormalizes new divergencies arising from string loops contribution
\cite{loops}. Considering this fact we can write our final
equations arising from the Weyl invariance of string theory in the
form:
\begin{eqnarray}
&& \nabla^2 H_{\mu\nu}
+ R_\mu{}^\alpha{}_\nu{}^\beta H_{\alpha\beta}
- R^\alpha{}_\mu H_{\alpha\nu}
- R^\alpha{}_\nu H_{\alpha\mu}
- \frac{1}{\alpha'} H_{\mu\nu} + O(\alpha') = 0 {,}
\nonumber\\&&
\nabla^\alpha H_{\alpha\nu} + O(\alpha') =0 {,} \qquad
H^\mu{}_\mu +O(\alpha') =0 {,}
\nonumber\\&&
R_{\mu\nu} +O(\alpha')= T^H_{\mu\nu} - \frac{1}{D-2}
G_{\mu\nu} T^H{}^\alpha{}_\alpha
\label{final}
 \end{eqnarray}
This is a system of consistent equations of motion derived in the
lowest order but if one wants to determine whether they can be
deduced from an effective lagrangian (and to find this lagrangian)
then one-loop contributions present in (\ref{final}) are not
sufficient. Comparing the effective equations of motion
(\ref{E},\ref{final}) with the general form of equations for spin 2
field discussed above we see that coefficients $E^{(i)}$
arising in the condition of string Weyl invariance are not equations
directly following from a lagrangian (\ref{eqhigh}) but some
combinations of them analogous to (\ref{finhigh}). In order to reverse
the procedure of passing from (\ref{eqhigh}) to (\ref{finhigh}) one
would need next to leading contributions in the conditions for
$\nabla^\mu H_{\mu\nu}$ and $H^\mu{}_\mu$ (\ref{final}).

Anyway the specific values of coefficients $a_i$ in the
effective action (\ref{higher}) are not important due to possibility
of field redefinition (\ref{redef}).
In string theory this possibility can be also interpreted as finite
renormalization of the field $H_{\mu\nu}$. Our result (\ref{final})
has coefficients specific for the minimal subtraction scheme which we
used performing the renormalization of background fields and
composite operators. Specific values of these coefficients are scheme
dependent. In case of arbitrary renormalization scheme the first
equation in (\ref{final}) will take the form:
\begin{eqnarray}
&& \nabla^2 H_{\mu\nu}
+ c_1 R_\mu{}^\alpha{}_\nu{}^\beta H_{\alpha\beta}
+ c_2 R^\alpha{}_\mu H_{\alpha\nu}
+ c_2 R^\alpha{}_\nu H_{\alpha\mu}
+ c_3 R H_{\mu\nu}
+ c_4 G_{\mu\nu } R^{\alpha\beta} H_{\alpha\beta}
\nonumber\\&&\qquad\qquad
- \frac{1}{\alpha'} H_{\mu\nu} + O(\alpha') = 0  {.}
\end{eqnarray}
$c_i$ are coefficients depending on the renormalization scheme and in
this order in $\alpha'$ all of them can be made arbitrary by the
field redefinition.

Let us summarize the obtained results.  We investigated the problem
of consistency of the equations of motion for spin-2 massive field in
curved spacetime and found that two different description
of this field are possible. First, for specific gravitational
background satisfying (\ref{restr}) one can build an action leading
to consistent equations including the tracelessness  and
transversality conditions. Another possibility (naturally arising
in string theory) consists in building the theory as
perturbation series in inverse mass. In the lowest order no
consistency problems arise and equations of motion have the form
(\ref{finhigh}).

Then we calculated the equations for the massive
spin-2 background field arising in sigma model approach to string
theory from the condition of quantum Weyl invariance in the lowest
order in $\alpha'$. The explicit form of the derived equations
(\ref{final}) appears to be a particular case of the
general equations in field theory (\ref{finhigh}). General agreement
between string theory and field theory is achieved if one takes
into account a possibility of making field redefinitions in the
effective action which in string theory calculations corresponds to
possibility of making finite renormalization of background fields.
We expect that in general in each order in $\alpha'$ the situation
remains the same and it is possible to construct the part of string
effective action quadratic in massive background field which should
lead to generalized mass-shell and transversality conditions.

To determine this part of the bosonic string effective action
completely one should also consider other massless background fields
including the dilaton. This will require investigation of strings
with curved world sheets and with non-vanishing extrinsic curvature
on the boundary which complicates the sigma-model calculations. From
the field theoretical point of view this will require to generalize
the analysis of consistency for the massive spin 2 field propagating
on background of both gravity and massless scalar field. We leave it
for a future publication.

\medskip

{\em Acknowledgements.}
This work was supported by GRACENAS grant, project 97-6.2-34; RFBR
grant, project 99-02-16617 and RFBR-DFG grant, project 96-02-00180.

%%%%%%%%%%%%%%%%%%%%%%%%%%%%%%%%%%%%%%%%%%%%%%%%%%%%%%%%%%%%%%%

\end{document}